\documentstyle[prl,pra,multicol,aps,epsf]{revtex}

\draft
\begin{document}
\draft
\pagenumbering{roma}
\title{Dynamics of entanglement  for coherent excitonic states
 in a system of two coupled quantum dots and cavity QED }
\author{Yu-xi Liu$^{a}$, \c{S}ahin Kaya \"Ozdemir$^{a,b}$, Masato Koashi$^{a,b}$, and  Nobuyuki Imoto$^{a,b}$}
\address{(a)The Graduate University for Advanced Studies (SOKEN),
  Hayama,Kanagawa, 240-0193,Japan }
\address{(b) CREST Research Team for Interacting Carrier Electronic, Hayama, Kanagawa, 240-0193, Japan}
\maketitle
\vspace{1mm}
\widetext
\begin{abstract}
The dynamics of the entanglement for coherent excitonic states in the system of two coupled large
semiconductor quantum dots ($R/a_{B}\gg 1$) mediated by a single-mode cavity field is investigated. Maximally
entangled coherent excitonic states can be generated by cavity field initially prepared in odd coherent
state. The entanglement of the excitonic coherent states between two dots reaches maximum when no photon is
detected in the cavity. The effects of the zero-temperature environment on the entanglement of excitonic
coherent state are also studied using the concurrence for two subsystems of the excitons.

\end{abstract}
\widetext

\begin{multicols}{2}
\pagenumbering{arabic}
\section{introduction}

Demonstration of the existence of fast quantum algorithms~\cite{shor} excited scientists to pour tremendous
enthusiasm into the quantum computation and information theory. In this relatively new area, scientists not
only make abstract theory, but also make efforts to find some physical systems to test and realize their
ideas. A representative list includes the trapped ions~\cite{cirac}, the liquid-state nuclear magnetic
resonance~\cite{chuang} and the solid-state material~\cite{haken}.

Nanotechnology opens technological possibilities to fabricate mesoscopic devices. Semiconductor
nanostructures, especially quantum dot structures, are very promising for the realization of quantum
computation and the quantum information processing. The quantum gate realization using quantum dots has been
proposed ~\cite{dot1}. Sanders et al. have discussed the scalable solid-state quantum computer based on
electric dipole transitions within coupled single-electron quantum dots~\cite{sanders}. Imamoglu et al.
propose  a scheme which realizes controllable interactions between two distant quantum dot spins by combining
the cavity quantum electrodynamics (QED) with electronic spin degrees in quantum dots~\cite{imamoglu}.
Entanglement of the exciton states in a single quantum dot or in a quantum dot molecule has been also
demonstrated experimentally~\cite{5,6}. The reference~\cite{7} investigate the entanglement of excitonic
states in  the system of the optically driven coupled quantum dots theoretically  and propose a method to
prepare maximally entangled Bell and Greenberger-Horne-Zeilinger states. Berman et al. has investigated the
dynamics of entangled states and a quantum control-not gate for an ensemble of four-spin molecules and shown
that entangled states can be generated using a resonant interaction between the spin system and an
electromagnetic $\pi$ pulse ~\cite{gp}.

The main focus of the references cited above is the entanglement of the excitonic orthogonal states. In fact,
not only the entangled orthogonal states but also  entangled nonorthogonal states play an important role in
quantum cryptography~\cite{fuchs} and quantum teleportation~\cite{wang}. One kind of entangled nonorthogonal
states is the entangled coherent states~\cite{bc,sj} which can be generated by a nonlinear Mach-Zender
interferometer. However up to now, no scheme for generating entangled excitonic coherent states has been
proposed. This paper will focus on this topic, especially, on the dynamics of the entangled excitonic
coherent states in the system of two large semiconductor quantum dots coupled by a single-mode field. The
conditions for the preparation of maximally entangled excitonic states are derived assuming that the
effective radius $R$ of each quantum dot is much larger than the Bohr-radius $a_{B}$ of the bulk-exciton. The
interactions between two quantum dots is mediated only by the single-mode cavity field.

This paper is organized as follows:  In Sec. II, we will describe the Hamiltonian of the system of interest,
and obtain the solutions of the operators which present dynamical quantity. Then we will give the wave
function of the system with certain initial states. In Sec. III, we will discuss the entanglement of the
excitonic coherent states in the system of the two coupled quantum dots by virtue of the
concurrence~\cite{wt,wt1}. The effects of environment on concurrence will be discussed in Sec. IV, and
finally the conclusion of this study will be presented in Sec. V.

\section{Hamiltonian and solutions}
The model that is analyzed in this study consists of two quantum dots which are placed into a single-mode
cavity. The quantum dots have large sizes satisfying the condition $R \gg a_{B}$. We also assume that there
are a few electrons excited from valence-band to conduction-band. Then the excitation density of the
Coulomb-correlated electron-hole pairs, excitons, in the ground state for each quantum dot is low. This, in
turn, implies that the average number of excitons is no more than one for an effective area of the excitonic
Bohr radius. Therefore  exciton operators can be approximated with boson operators, and all nonlinear terms
including exciton-exciton interactions and the phase space filling effect can be neglected. The ground energy
of the excitons in each dots is assumed to be the same. There is a resonant interaction between the
single-mode cavity field and the excitons. The distance between two quantum dots is assumed to be much larger
than the optical wavelength $\lambda$ of the cavity field, so the interaction between quantum dots can be
neglected. Under the rotating wave approximation, the Hamiltonian for this system can be written
as~\cite{Hui}
\begin{equation}
H_{0}=\hbar\omega\hat{a}^{\dagger}\hat{a}+
\hbar\omega\sum_{m=1}^{2}\hat{b}^{\dagger}_{m}\hat{b}_{m}+
\hbar\sum_{m=1}^{2}g_{m}(\hat{b}^{\dagger}_{m}
\hat{a}+\hat{a}^{\dagger}\hat{b}_{m}) \label{eq:h0}
\end{equation}
where $\hat{a}^{\dagger}(\hat{a})$ are the operators of the cavity
field with frequency $\omega$,
$\hat{b}^{\dagger}_{m}(\hat{b}_{m})$ denote the exciton operators
with the same frequency $\omega$ of the cavity field, and $m=1(2)$
represent the first dot or the second dot. The coupling constants
between quantum dot one (two) and cavity field are represented by
$g_{m}$ with $m=1, 2$. Without loss of generality, we can take the
two coupling constants between the cavity field and the quantum
dots as different. The Heisenberg equations of motion for the
operators of the cavity field $\hat{a}(\hat{a}^{\dagger})$ and the
excitons $\hat{b}_{m}(\hat{b}^{\dagger}_{m})$ can be easily
obtained as
\begin{mathletters}
\begin{eqnarray}
\frac{\partial \hat{a}}{\partial t}&=&-i\omega \hat{a}-i g_{1}
\hat{b}_{1}-i g_{2}\hat{b}_{2} \\
\frac{\partial \hat{b}_{1}}{\partial t}&=&-i\omega \hat{b}_{1}-i
g_{1}\hat{a}\\
\frac{\partial \hat{b}_{2}}{\partial t}&=&-i\omega \hat{b}_{2}-i
g_{2}\hat{a}
\end{eqnarray}
\end{mathletters}
It is very easy to obtain the solutions of the above operator
equations as
\begin{mathletters}
\begin{eqnarray}
&&\hat{a}(t)=\cos (gt)\hat{a}(0) e^{-i\omega t}
-i\frac{g_{1}}{g}\sin (gt)\hat{b}_{1}(0) e^{-i\omega t}\nonumber\\
&&-i\frac{g_{2}}{g}\sin(gt)\hat{b}_{2}(0)e^{-i\omega t}\\
&&\hat{b}_{1}(t)=-i\frac{g_{1}}{g}\sin (gt)\hat{a}(0) e^{-i\omega
t} +\frac{g_{1}^2 \cos(gt)+g_{2}^2}{g^2}\hat{b}_{1}(0)
e^{-i\omega t}\nonumber\\
&&+\frac{g_{1}g_{2}[\cos(gt)-1]}{g^2}\hat{b}_{2}(0)e^{-i\omega t}\\
&&\hat{b}_{2}(t)=-i\frac{g_{2}}{g}\sin (gt)\hat{a}(0) e^{-i\omega
t}+\frac{g_{1}g_{2}[\cos(gt)-1]}{g^2}\hat{b}_{1}(0)e^{-i\omega t}\nonumber\\
&&+\frac{g_{2}^2 \cos(gt)+g_{1}^2}{g^2}\hat{b}_{2}(0) e^{-i\omega
t}
\end{eqnarray}
\end{mathletters}
with $g=\sqrt{g_{1}^2+g_{2}^2}$.

We assume that the cavity field is initially prepared in a
 superposition of two distinct coherent states $|\alpha_{1}\rangle$ and
$|\alpha_{2}\rangle$ in the normalized form of
\begin{equation}
|\psi(0)\rangle_{c}=C|\alpha_{1}\rangle+D|\alpha_{2}\rangle.
\label{eq:op}
\end{equation}
The two modes of excitons are in the vacuum states
$|0\rangle_{1}|0\rangle_{2}$ with label $1(2)$ denoting exciton
mode one (two). The whole initial state for the excitons and the
cavity field can be expressed as
\begin{equation}
|\Psi(0)\rangle=(C|\alpha_{1}\rangle+D|\alpha_{2}\rangle)
\otimes |0\rangle_{1}|0\rangle_{2}.
\end{equation}
The time dependent wave function $|\Psi(t)\rangle$ of the whole
system can use the time evolution operator $U(t)=e^{-i(Ht/\hbar)}$
and initial state $|\Psi(0)\rangle$ to express as
$|\Psi(t)\rangle=U(t)|\Psi(0)\rangle$, that is
\begin{equation}
|\Psi(t)\rangle=U(t)(C|\alpha_{1}\rangle+
D|\alpha_{2}\rangle)\otimes |0\rangle_{1}|0\rangle_{2}.
\label{eq:1}
\end{equation}
According to the definition of the coherent state, any coherent
state of the cavity field  has  the following form
\begin{equation}
|\alpha\rangle=\exp[\alpha \hat{a}^{\dagger}(0)-\alpha^{*} \hat{a}(0)]
|0\rangle
\end{equation}
where ($\ast$) denotes complex conjugation. Then (\ref{eq:1}) can be rewritten as
\begin{eqnarray}
|\Psi(t)\rangle&=&U(t)[C \exp(\alpha_{1} \hat{a}^{\dagger}(0)-\alpha_{1}^{*}
\hat{a}(0)) \nonumber \\
&+&D \exp(\alpha_{2} \hat{a}^{\dagger}(0)-\alpha_{2}^{*} \hat{a}(0))]
|0\rangle |0\rangle_{1}|0\rangle_{2}.
\label{eq:2}
\end{eqnarray}
We may interpolate unit operator $U^{\dagger}(t)U(t)$ into
Eq.(\ref{eq:2}) and consider the properties of the time evolution
operator $U^{\dagger}(t)OU(t)=O(t)$ and $U(t)|0\rangle=|0\rangle$,
then Eq.(\ref{eq:2}) becomes as
\begin{eqnarray}
|\Psi(t)\rangle&=&C|\alpha_{1} u(t)\rangle \otimes|\alpha_{1}v_{1}(t)
\rangle_{1}
\otimes|\alpha_{1}v_{2}(t)\rangle_{2} \nonumber \\
&+&
D|\alpha_{2} u(t)\rangle \otimes |\alpha_{2}v_{1}(t)\rangle_{1}
\otimes|\alpha_{2}v_{2}(t)\rangle_{2}
\end{eqnarray}
with
\begin{mathletters}
\begin{eqnarray}
u(t)&=&\cos (gt) e^{-i\omega t},\\
v_{1}(t)&=&-i\frac{g_{1}}{g}\sin (gt)e^{-i\omega t},\\
v_{2}(t)&=&-i\frac{g_{2}}{g}\sin(gt)e^{-i\omega t}.
\end{eqnarray}
\end{mathletters}
For convenience, we will drop the time dependence $t$  from
$u(t)$, $v_{1}(t)$, and $v_{2}(t)$ in the following expressions.
It is very clear that three subsystems become entangled with the
time evolution except at some critical times $t=n\pi/2g$ for
integer $n$. But what we are interested in is the entanglement
between two subsystems of excitons, so we trace over mode of the
cavity field in the density matrix of the system given as
$|\Psi(t)\rangle\langle\Psi(t)|$ to obtain the reduced density
matrix $\rho(t)={\rm Tr}(|\Psi(t)\rangle\langle\Psi(t)|)$ of
excitons for any arbitrary time $t$. This partial trace operation
enables us to describe the dynamics of the excitons in the coupled
quantum dot system without any reference to the cavity field. Then
the reduced density matrix $\rho(t)$ is found as
\begin{eqnarray}
\rho(t)&=&[C^{*}D {\cal N} (|\alpha_{2}v_{1}\rangle\langle\alpha_{1}v_{1}|) _{1}
\otimes(|\alpha_{2}v_{2}\rangle\langle\alpha_{1}v_{2}|)_{2}+h.c.] \nonumber \\
&+&|C|^2 (|\alpha_{1}v_{1}\rangle\langle\alpha_{1}v_{1}|) _{1}
\otimes(|\alpha_{1}v_{2}\rangle\langle\alpha_{1}v_{2}|)_{2}
\nonumber\\
&+&|D|^2 (|\alpha_{2}v_{1}\rangle\langle\alpha_{2}v_{1}|) _{1}
\otimes(|\alpha_{2}v_{2}\rangle\langle\alpha_{2}v_{2}|)_{2}.
\label{eq:p}
\end{eqnarray}
Here $h.c.$ denotes complex conjugate. ${\cal N}$  is an inner product of two coherent states $|\alpha_{2}
u\rangle$ and $|\alpha_{1} u\rangle$, and ${\cal N}=\exp\left[-\cos^{2}(gt)(\frac{1}{2}|\alpha_{1}|^2+
\frac{1}{2}|\alpha_{2}|^2-\alpha^{*}_{1}\alpha_{2} )\right]$. It is evident that unitary evolution of the
wave function cannot change its purity, however tracing out the cavity field reduces the other subsystems
into a statistical mixture.

\section{entanglement of excitonic states}

Concurrence is a useful tool to measure entanglement between two
subsystems of $2 \times 2$ biparte mixed and pure states.
According to the references~\cite{wt,wt1}, we assume that $1$ and
$2$ is a pair of qubits, and the density matrix of the pair is
$\rho_{12}$. Then the concurrence of the density matrix
$\rho_{12}$ is defined as:
\begin{equation}
{\cal C}_{12}=max \{ \lambda_{1}-\lambda_{2}-\lambda_{3}-\lambda_{4},0 \}
\label{eq:c}
\end{equation}
where $\lambda_{1}$, $\lambda_{2}$, $\lambda_{3}$, and
$\lambda_{4}$ given in decreasing order are the square roots of
eigenvalues for matrices
\begin{equation}
M_{12}=\rho_{12}(\sigma_{1y}\otimes \sigma_{2y})
\rho^{*}_{12}(\sigma_{1y}\otimes \sigma_{2y})
\label{eq:m}
\end{equation}
with Pauli matrix
\begin{equation}
\sigma_{1y}=\left(\begin{array}{cc} 0 &-i \\ i & 0\end{array}\right);
\hspace{0.5cm}
\sigma_{2y}=\left(\begin{array}{cc} 0 &-i \\ i & 0\end{array}\right),
\end{equation}
and ($\ast$) denoting complex conjugation in the standard basis $\{|00\rangle, |01\rangle, |10\rangle,
|11\rangle \}$, and $\sigma_{1y}$  and $\sigma_{2y}$ are expressed in the same basis. The entanglement of
formation is a monotonically increasing function of ${\cal C}_{12}$.  ${\cal C}_{12}=0$ corresponds to an
unentangled state, and ${\cal C}_{12}=1$ corresponds to a maximally entangled state.

First, we make a suitable transformation on Eq.(\ref{eq:p}) so that we can define the qubit and discuss the
entanglement between two modes of excitons using the concurrence. We can choose two different coherent
states, which are nonorthogonal and form a super-complete set, as a basis to span a two-dimensional Hilbert
space. But it is not suitable to define qubits using two-dimensional super-complete set of basis. However,
one can always rebuild two orthogonal and normalized states as basis of the two-dimensional Hilbert space
using original two coherent states~\cite{dh}. In our system, we can choose two time-dependent excitonic
coherent states to define qubits. We define the zero qubit state as
$|0\rangle_{m}=|\alpha_{1}v_{m}(t)\rangle$; and one qubit state as
$|1\rangle_{m}=(|\alpha_{2}v_{m}(t)\rangle- \langle\alpha_{1}v_{m}(t)|\alpha_{2}v_{m}(t)\rangle
|\alpha_{1}v_{m}(t)\rangle)/N_{m}(t)$ with
$N_{m}(t)=\sqrt{1-|\langle\alpha_{1}v_{m}(t)|\alpha_{2}v_{m}(t)\rangle|^2} =\sqrt{1-P_{m}^{2}(t)}$ for
exciton systems of mode one $m=1$ and two $m=2$. The above definition for qubits is equivalent to the
definition in reference~\cite{kim}. In this new basis, the reduced density matrix (\ref{eq:p}) can be
rewritten into
\begin{eqnarray}
\rho&=&\{ C^{*}D {\cal N}[(N_{1}(t)|1\rangle\langle 0|)_{1}+
P_{1}(t)(|0\rangle\langle 0|)_{1}]\otimes \nonumber \\
&&[(N_{2}(t)|1\rangle\langle 0|)_{2}+P_{2}(t)
(|0\rangle\langle 0|)_{2}] +h.c.\} \nonumber\\
&&|D|^{2}[(N_{1}(t)|1\rangle+P_{1}(t)|0 \rangle)
(N^{*}_{1}(t)\langle 1|+P_{1}^{*}(t)\langle 0|)]_{1}
\otimes\nonumber\\
&&[(N_{2}(t)| 1\rangle+P_{2}(t)|0 \rangle)
(N^{*}_{2}(t)\langle 1|+P^{*}_{2}(t)\langle 0|)]_{2} \nonumber \\
&&+|C|^2(|0\rangle\langle 0|)_{1}\otimes
(|0 \rangle\langle 0|)_{2}.
\label{eq:ee}
\end{eqnarray}

For simplicity, we assume that two quantum dots are completely the same so that the interaction between
cavity field  and two quantum dots are equal to each other,  that is $g_{1}=g_{2}$. We consider the cases
where the cavity field is initially prepared in the even or odd coherent states. It has been shown that such
mesoscopic coherent superpositions can be generated by experimentalists using cavity QED~\cite{haroche} or a
trapped ion~\cite{win}. For the cavity field initially prepared in the odd coherent state, Eq.(\ref{eq:op})
can be written as
\begin{equation}
|odd\rangle=N_{-}(|\alpha\rangle-|-\alpha\rangle) \label{eq:odd}
\end{equation}
with normalization constant
$N_{-}=(2-2e^{-2|\alpha|^2})^{-\frac{1}{2}}$. Under the above
conditions, we can write $N_{1}(t)=N_{2}(t)=M=\sqrt{1-P^2(t)}$ and
$P=P_{1}(t)=P_{2}(t)=e^{-\sin^2(gt)|\alpha|^2 }$. Then the reduced
density operator (\ref{eq:ee}) can be rewritten in the matrix
representation as follows
\end{multicols}
\widetext
\begin{equation}
R(\rho)=\frac{1}{2(1-e^{-2|\alpha|^2})} \left( \begin{array}{cccc}
1-2{\cal N}P^2+P^4& -{\cal N}MP+MP^3&
-{\cal N}MP+MP^3& -{\cal N}M^2+M^2P^2 \\
-{\cal N}MP+MP^3&  M^2P^2 & M^2P^2 & M^3P\\
-{\cal N}MP+MP^3&  M^2P^2 & M^2P^2 & M^3P\\
-{\cal N}M^2+M^2P^2& M^3P& M^3P& M^4
 \end{array}\right)
\label{eq:r}
\end{equation}
\widetext
\begin{multicols}{2}
\noindent in the standard basis $\{|00\rangle, |01\rangle,
|10\rangle, |11\rangle \}$ where every basis vector such as
$|00\rangle$ denotes state $|0\rangle_{1}\otimes|0\rangle_{2}$. We
can obtain the square roots of eigenvalues of the matrix
(\ref{eq:m}) corresponding to the matrix (\ref{eq:r}) as
\begin{mathletters}
\begin{eqnarray}
\lambda_{1}&=&\frac{(1-e^{-2\sin^{2}(gt)|\alpha|^2})
(1+e^{-2\cos^{2}(gt)|\alpha|^2})}
{2(1-e^{-2|\alpha|^2})}\\
\lambda_{2}&=&\frac{(1-e^{-2\sin^{2}(gt)|\alpha|^2})
(1-e^{-2\cos^{2}(gt)|\alpha|^2})}
{2(1-e^{-2|\alpha|^2})}\\
\lambda_{3}&=&\lambda_{4}=0.
\end{eqnarray}
\end{mathletters}
Then the  concurrence $C_{o}$ of the prepared state with an
initial odd coherent state of the cavity field is found as
\begin{equation}
C_{o}=\frac{e^{-2\cos^{2}(gt)|\alpha|^2}(1-e^{-2\sin^{2}(gt)|\alpha|^2})}
{(1-e^{-2|\alpha|^2})}.
\label{eq:ww}
\end{equation}

\begin{figure}[h]
\epsfxsize=5cm \centerline{\epsffile{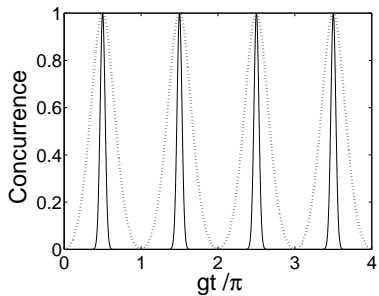}} \vspace{2mm} \caption[]{Time evolution of the concurrence
$C_{o}$ is plotted as a function of $\frac{gt}{\pi}$ for (a) $|\alpha|=1$ (dashed), (b) $|\alpha|=5$
(solid).}\label{fig1}
\end{figure}

We can also obtain the average photon number $\bar{n}$ of the
cavity field as follows
\begin{equation}
\bar{n}=\langle a^{\dagger}a \rangle=\frac{|\alpha|^2
\cos^2(gt)(1+e^{-2|\alpha|^2})}{1-e^{-2|\alpha|^2}}\label{eq:20}.
\end{equation}
 Eqs.(\ref{eq:ww}-\ref{eq:20}) show that two subsystems of the excitons
becomes maximally entangled for any time satisfying the condition $gt=(n+\frac{1}{2})\pi$ for any intensity
$|\alpha|^2$ of the coherent cavity field, and in this moment no photon of the cavity field can be detected.
Fig. {\ref{fig1}} shows the time evolution of (\ref{eq:ww}) for two different intensity values $|\alpha|^2$
of the cavity field. It is clearly seen that maximally entangled coherent states of excitons can be prepared
by the cavity field of odd coherent states. The entanglement periodically  reaches its maximum. It is also
found that when the excitonic states are in a maximally entangled state, the average photon number of the
cavity field is zero (see Fig.\ref{fig3a}). This enables us to monitor the status of the system during
preparation of the entangled excitonic coherent states by detecting the photon number of the cavity field.

If the cavity field is initially prepared in an even coherent state
\begin{equation}
|even\rangle=N_{+}(|\alpha\rangle+|-\alpha\rangle)
\end{equation}
where the normalization constant is
$N_{+}=(2+2e^{-2|\alpha|^2})^{-\frac{1}{2}}$, using the same
procedure the concurrence $C_{e}$ is found as
\begin{equation}
C_{e}=\frac{e^{-2\cos^{2}(gt)|\alpha|^2}(1-e^{-2\sin^{2}(gt)|\alpha|^2})}
{(1+e^{-2|\alpha|^2})}, \label{eq:ce}
\end{equation}
and the average number $\bar{n}$ of the cavity field is
\begin{equation}
\bar{n}=\langle a^{\dagger}a\rangle=\frac{
|\alpha|^2\cos^2(gt)(1-e^{-2|\alpha|^2})}{1+e^{-2|\alpha|^2}}.
\end{equation}
\noindent From Eq.(\ref{eq:ce}), it is found that we can not obtain maximally entangled states when the
cavity field is initially in the even coherent state. But the analytical expression (\ref{eq:ce}) shows that
when the coherent intensity $|\alpha|$ of the cavity field tends to infinity, we can approximately obtain
maximally entangled excitonic coherent state at

\begin{figure}[h]
\epsfxsize=5cm \centerline{\epsffile{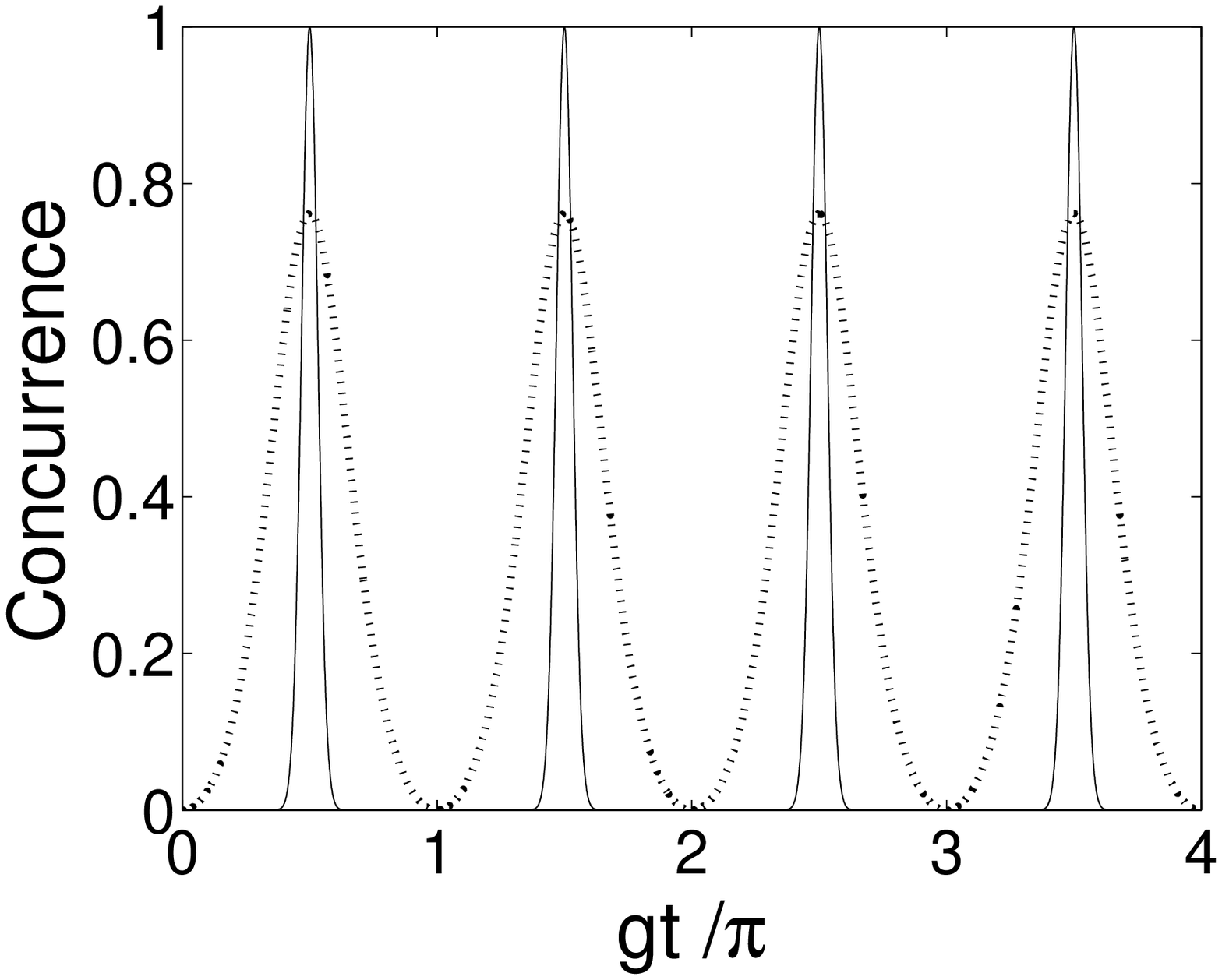}}\vspace{3mm} \caption[]{Time evolution of the concurrence
$C_{e}$ is plotted as a function of $gt/\pi$ for (a) $|\alpha|=1$ (dashed), (b) $|\alpha|=5$
(solid).}\label{fig2}
\end{figure}
\vspace{-3.5mm}
\begin{figure}[h]
\hspace*{-6mm} \epsfxsize=4.5cm \epsfbox{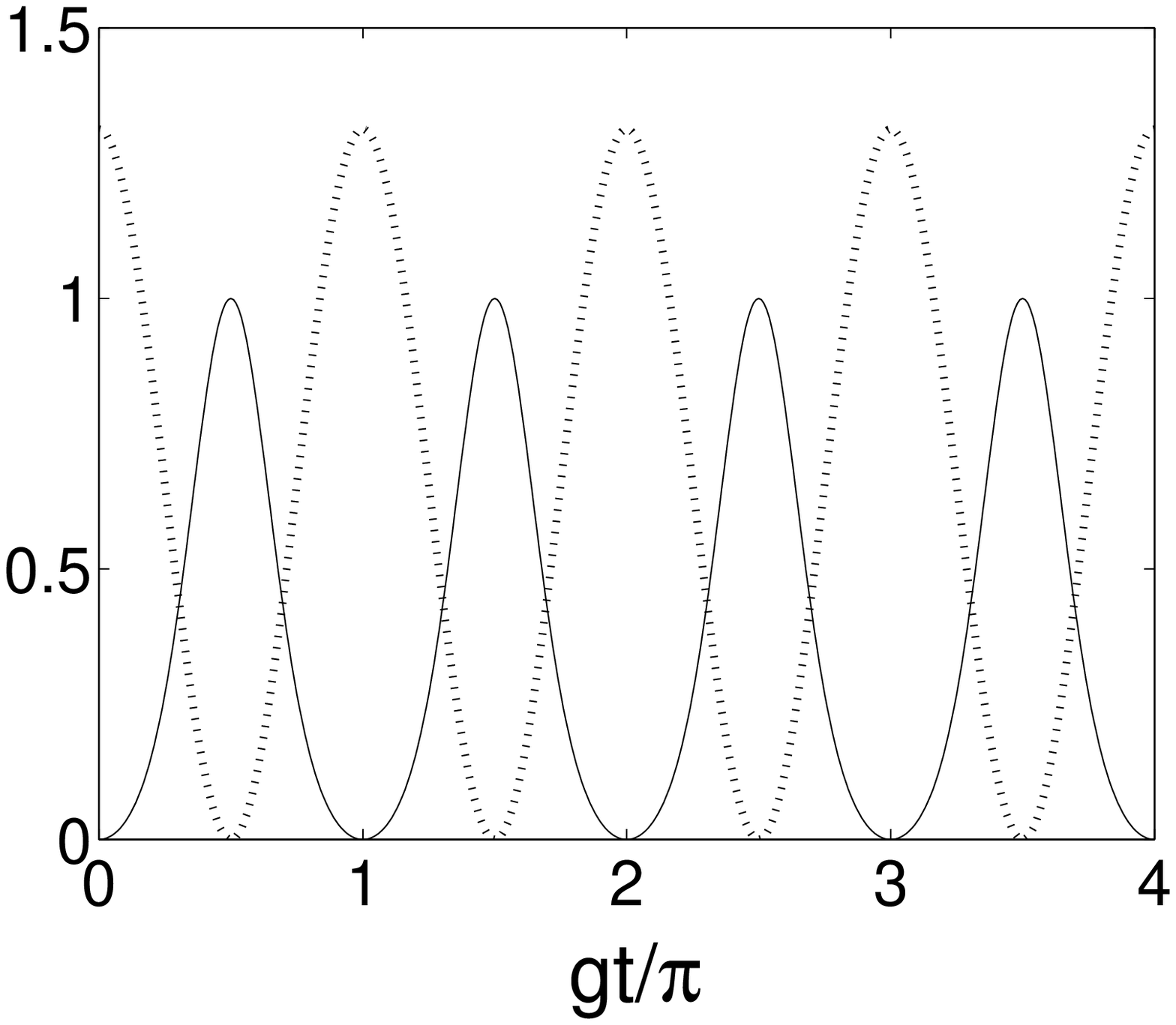}\hspace*{0mm}\epsfxsize=4.25cm
\epsfbox{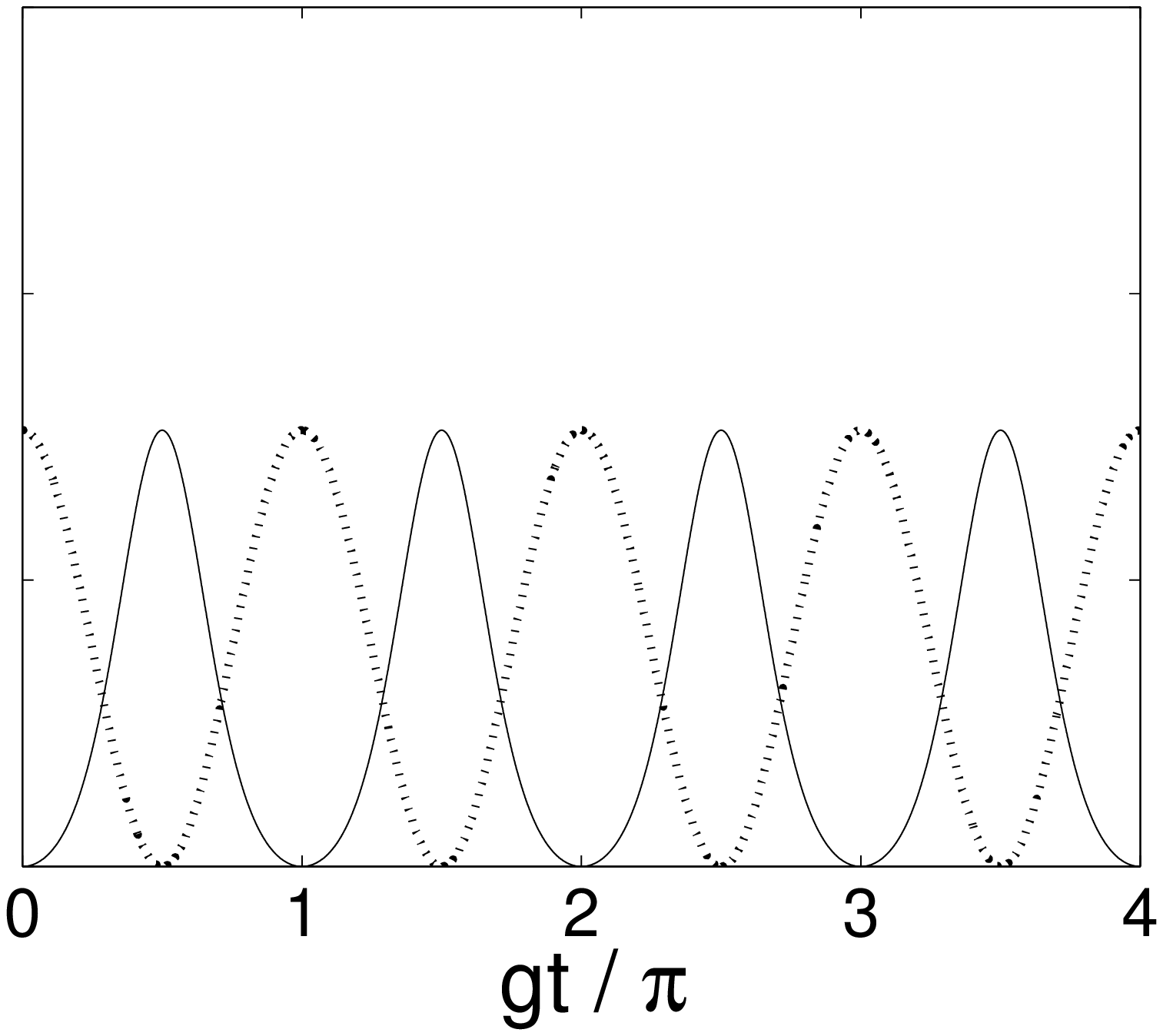}\vspace{3mm} \caption[]{Time evolutions of the concurrence (solid curve) and average
photon number $\bar{n}$ (dashed curve) are plotted as a function of $gt/\pi$ for cavity field initially
prepared with $|\alpha|=1$ as odd coherent state (left) and even coherent state (right).}\label{fig3a}
\end{figure}
\noindent  time $gt=(n+\frac{1}{2})\pi$. In fact numerical results show that when the coherent intensity
$|\alpha|\ge 2$, the concurrence approximately reaches its maximum at the same time $gt=(n+\frac{1}{2})\pi$.
Fig.\ref{fig2} shows the time evolution of the concurrence $C_{e}$ for two different $|\alpha|$  values of
the cavity field. It is concluded that a quasi-maximally entangled excitonic coherent states can be prepared
using the initial even coherent state of the cavity field when the coherent intensity $|\alpha|$ becomes
intensive.

In Fig.\ref{fig3a}, a comparison of the degree of entanglement and
its relation with the average number of photons ($\bar{n}$) in the
cavity for cavity fields initially prepared in odd and even
coherent states is presented. It is seen that (i) Both the
concurrence and $\bar{n}$ shows a periodic evolution,
(ii)Concurrence reaches to its maximum value when $\bar{n}=0$ for
both cases, and (iii) for the cavity field initially prepared in
odd coherent state, it is possible to obtain maximally entangled
states for some evolution times, but this is not true for cavity
field initially prepared in even coherent state.

\section{effect of environment on the entanglement of excitons}
In this section, we will focus our attention on how the environment with zero temperature affects the
entanglement of the excitons. We assume that the energy of the system is dissipated only by the interaction
of the excitons in the quantum dots with the multi-modes of electromagnetic radiation. Our discussion is
limited to these two completely equivalent quantum dots, that is, they have the same coupling constants with
the cavity field, and also have the same dissipative dynamics. Under above considerations, we have the
following Hamiltonian with the rotating-wave approximation as
\begin{eqnarray}
H&=&H_{0}+\hbar\sum_{m=1}^{2}\sum_{k}\omega_{m,k}a^{\dagger}_{m,k}a_{m,k}
\nonumber \\
&+&\sum_{m=1}^{2}\sum_{m,k}\kappa_{m,k}
(a_{m,k}b^{\dagger}_{m}+a^{\dagger}_{m,k}b_{m})
\end{eqnarray}
where $H_{0}$ is determined by (\ref{eq:h0}), and $a^{\dagger}_{m,k}(a_{m,k})$ denote the multi-modes
electromagnetic field operators with the frequencies $\omega_{m,k}$, and $m=1(2)$ means the quantum dot one
(two). Using the assumption $g_{1}=g_{2}=\kappa=g/\sqrt{2}$, and $\kappa_{1,k}=\kappa_{2,k}$, we can write
the equations of motion for all operators as follows
\begin{mathletters}
\begin{eqnarray}
&&\frac{\partial \hat{a}}{\partial t}=-i\omega_{0} \hat{a}-i \kappa
\hat{b}_{1}-i \kappa\hat{b}_{2},  \label{eq:q1}\\
&&\frac{\partial \hat{b}_{m}}{\partial t}=-i\omega_{0} \hat{b}_{m}-i
\kappa\hat{a}-i\sum_{k}\kappa_{m,k}a_{m,k},\\
&&\frac{\partial a_{m,k}}{\partial t}=-i\omega_{m,k}a_{m,k}-i\kappa_{m,k}b_{m}.
\label{eq:qq}
\end{eqnarray}
\end{mathletters}
We know that only the solution of the cavity field operator is enough to solve our problem. We apply the
Laplace transformation and the Wigner-Weisskopf approximation~\cite{lou} to the above set of
Eqs.(\ref{eq:q1}-\ref{eq:qq}). So the time-dependent solution of the cavity field operator with the
zero-temperature environment can be obtained as
\begin{eqnarray}
a(t)&=&u^{\prime}(t)a(0)+v^{\prime}_{1}(t)b_{1}(0)+v^{\prime}_{2}(t)b_{2}(0)
\nonumber\\
&+&\sum_{k}v_{1,k}(t)a_{1,k}(0)+\sum_{k}v_{2,k}(t)a_{2,k}(0)
\end{eqnarray}
where
\begin{mathletters}
\begin{eqnarray}
u^{\prime}(t)&=&\frac{e^{-\frac{\gamma}{4}t}
[\gamma(e^{\frac{i}{4}\Delta t}-e^{-\frac{i}{4}\Delta t})
+i\Delta(e^{\frac{i}{4}\Delta t}+e^{-\frac{i}{4}\Delta
  t})]e^{-i\omega t}}{i2\Delta} \label{eq:sh1}\\
v_{1}^{\prime}(t)&=&v_{2}^{\prime}(t)=
\frac{2\kappa[e^{\frac{1}{4}(-\gamma+i\Delta)t}-
e^{\frac{1}{4}(-\gamma-i\Delta)t}] e^{-i\omega t}}{i\Delta}
\label{eq:sh2}
\end{eqnarray}
\end{mathletters}
with $\Delta=\sqrt{32 \kappa^2-\gamma^{2}}$, and $\gamma=2\pi
\epsilon(\omega_{0})|\kappa(\omega_{0})|^2$ where
$\epsilon(\omega)$ is a distribution function of the multimode
electromagnetic radiation field. A small Lamb frequency shift has
been neglected in above Eqs.(\ref{eq:sh1}-\ref{eq:sh2}). We take
the same steps as in the Sect. II, and obtain the time-dependent
wave function of the whole system with cavity field initially in
the odd  mesoscopic coherent superposition of states
(\ref{eq:odd}) and other subsystems in the vacuum states
\begin{eqnarray}
&&|\Psi(t)\rangle= C |\alpha u^{\prime }(t)\rangle\otimes
\prod_{m=1}^{2}|\alpha v^{\prime }_{m}(t)\rangle
\otimes\prod_{m=1}^{2}\prod_{k}|\alpha v^{}_{m,k}(t)\rangle\nonumber\\
&&-C|-\alpha u^{\prime }(t)\rangle\otimes \prod_{m=1}^{2}|-\alpha
v^{\prime }_{m}(t)\rangle \otimes\prod_{m=1}^{2}\prod_{k}|-\alpha
v^{}_{m,k}(t)\rangle.
\end{eqnarray}
After tracing over the environment and the cavity field, we obtain
the reduced density matrix of  two subsystems of the excitons as
\begin{eqnarray}
\rho^{\prime}_{o}(t)&=&[|C|^2 {\cal N}^{\prime}
(|\alpha_{2}v^{\prime
}_{1}\rangle\langle\alpha^{\prime}_{1}v_{1}|) _{1}
\otimes(|\alpha_{2}v^{\prime }_{2}\rangle
\langle\alpha_{1}v^{\prime }_{2}|)_{2}
+h.c.] \nonumber \\
&+&|C|^2(|\alpha_{1}v^{\prime}_{1}\rangle
\langle\alpha_{1}v^{\prime}_{1}|)_{1}
\otimes(|\alpha_{1}v^{\prime}_{2}\rangle
\langle\alpha_{1}v^{\prime}_{2}|)_{2}
\nonumber\\
&+&|C|^2(|\alpha_{2}v^{\prime}_{1}\rangle
\langle\alpha_{2}v^{\prime}_{1}|)_{1}
\otimes(|\alpha_{2}v^{\prime}_{2}\rangle
\langle\alpha_{2}v^{\prime}_{2}|)_{2} \label{eq:mm}
\end{eqnarray}
with ${\cal N}^{\prime}=e^{-|\alpha|^2(1-2|v^{\prime}_{1}|^2)}$,
where the condition $\sum_{k}|v_{1,k}(t)|^2+\sum_{k}|v_{2,k}(t)|^2
=1-|u^{\prime}(t)|^2-|v^{\prime}_{1}(t)|^2-|v^{\prime}_{2}(t)|^2$,
which comes from the commutation relation
$[a(t),a^{\dagger}(t)]=1$, is used. Then we choose  the basis as
the section II, and use the same step to obtain the concurrence
corresponding to the reduced density matrix operator (\ref{eq:mm})
\begin{equation}
C^{\prime}_{o}=\frac{e^{-2|\alpha|^2(1-2|v^{\prime}_{1}|^2)} (1-e^{-4|\alpha|^2|v^{\prime}_{1}|^2})}
{(1-e^{-2|\alpha|^2})}, \label{eq:ww2}
\end{equation}
where $v^{\prime}_{1}$  is determined by (\ref{eq:sh2}). When $\gamma=0$ which means that the whole system
has no any dissipation of energy, then we find that (\ref{eq:ww2}) returns to ideal case (\ref{eq:ww}). Figs.
4 and 5 show that the entanglement of the two coherent excitonic states is gradually reduced by the
environment with time evolution. If the decay rate $\gamma$ of the excitons is fixed, then the higher the
intensity of the cavity field is, the faster the decay of entanglement is. In the same way, if the intensity
of the cavity field is fixed, then for a system with larger decay rate $\gamma$ of excitons, the amount of
entanglement reduces faster.
\begin{figure}[h]
\epsfxsize=5.5 cm \centerline{\epsffile{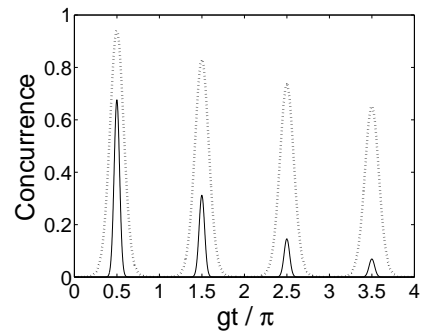}} \vspace{5.5mm}\caption[]{Time evolution of the concurrence
$C_{o}$ is plotted as a function of $gt/\pi$ for  $\gamma/g=0.01$ when $|\alpha|=5$ (solid), and $|\alpha|=2$
(dashed).}
\end{figure}
\begin{figure}
\epsfxsize=5.5cm \centerline{\epsffile{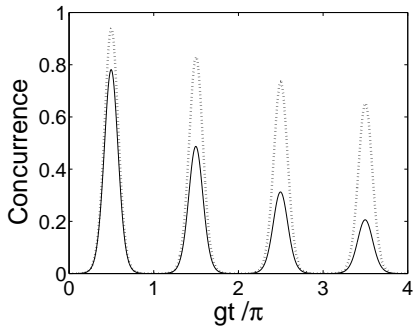}} \vspace{2mm} \caption[]{Time evolution of the concurrence
$C_{o}$ is plotted as a function of $gt/\pi$ for  $\gamma/g=0.01$ (dashed), and $\gamma/g=0.04$ (solid) when
$|\alpha|=2$.}
\end{figure}

\section{Conclusion}
The  entanglement of formation for coherent states of the excitons in the system of two coupled quantum dots
by cavity field is investigated. We find that a maximally entangled coherent state of two-modes excitons can
be prepared by an odd coherent state of the cavity field. The status of  this entangled excitonic coherent
state can be probed by the measurement of the cavity field. No photon detection in the cavity implies the
preparation of maximally entangled excitonic coherent states. In the case of photon detection of any number,
it can be said that the excitons are not  maximally entangled.

The effects of the zero-temperature environment on the entanglement of excitons is also studied by using
concurrence. It is found that the amount of entanglement between two excitonic coherent states is reduced by
the environment with time evolution due to the loss of photons of the cavity field to the environment. If
leakage from the cavity to the environment is monitored by photon counting detectors, an absence of
photo-detections implies that maximally entangled state of the excitons can be found in the coupled quantum
dot system with time evolution. In case of any photon detection, we are certain that the system can not reach
to a maximally entangled state. If the decay rate $\gamma$ of the excitons is given, for the system with
higher cavity field intensity, the decrease in the amount of entanglement is faster. It is also understood
that if the coherent intensity of the cavity field is given, then the large decay rate $\gamma$ of excitons
corresponds to a fast reduction in entanglement of the system.

To have an idea on the characteristic time-scales of the dynamics of entanglement in our model, we can take
$\hbar g\simeq 0.5~{\rm meV}$ which may be possible to obtain in a system of large area quantum dot with
cavity \cite{imamoglu}. In that case, for an ideal cavity with initial field prepared as an odd coherent
state,  the first maximally entangled state could be observed after a time evolution of $2~{\rm ps}$ and this
would repeat itself periodically with a period of $4~{\rm ps}$. For a practical cavity, the finite cavity
lifetime may be taken as $\gamma^{-1}\simeq 10 ~{\rm ps}$ \cite{imamoglu}. For these $g$ and $\gamma$ values,
we can find $\gamma/g\simeq 0.13$. With $|\alpha|=2$ ($|\alpha|=1$), the maximum value for concurrence would
be $\sim0.46$ ($\sim0.8$)which occurs at $t\simeq2~{\rm ps}$. After a time evolution of $4~{\rm ps}$, the
second peak for the concurrence would appear with a value of $\sim0.17$ ($\sim0.53$). It is clearly seen that
for a practical scheme the main parameters which would affect the dynamics of the system are the fast rate of
the photon loss from the cavity and intensity of the initial cavity field.

\section { Acknowledgments}
The authors thank to A. Miranowicz for helpful discussions. Y. L. is grateful  to the Japan Society for the
Promotion of Science (JSPS) for support. This work is also supported by Grant-in-Aid for Scientific Research
(B) (Grant No.~12440111) by Japan Society for the Promotion of Science.

\end{multicols}
\end{document}